\documentclass[11pt]{article}

\usepackage{amsmath, amssymb}
\usepackage{graphicx}
\usepackage{cite}
\usepackage{geometry}
\usepackage{tabularx}
\usepackage{booktabs}
\usepackage{caption}
\usepackage{subcaption}
\usepackage{hyperref}
\usepackage{xcolor}

\usepackage{setspace}
\title{AI-Based Detection of Temporal Changes in MR‑Linac Images Acquired During Routine Prostate Radiotherapy}

\author{
Seungbin Park, PhD$^{1}$\and
Peilin Wang, MS$^{1,2}$ \and
Ryan Pennell, MS$^{3}$ \and
Emily S. Weg, MD$^{3}$ \and
Himanshu Nagar, MD, MS$^{4}$ \and
Timothy McClure, MD$^{1}$ \and
Mert R. Sabuncu, PhD$^{1,5}$ \and
Daniel Margolis, MD$^{1}$ \and
Heejong Kim, PhD$^{1}$\thanks{Corresponding author} 
}

\date{}

\begin{document}

\maketitle

\vspace{-5mm}

\noindent
{\footnotesize
$^{1}$Radiology, Weill Cornell Medicine, New York, NY 10065, United States\\ 
$^{2}$Department of Health Technology and Informatics, The Hong Kong Polytechnic University, Hung Hom, Hong Kong\\
$^{3}$Radiation Oncology, Weill Cornell Medicine, New York, NY 10065, United States \\ 
$^{4}$Department of Radiation Oncology, Memorial Sloan Kettering Cancer Center, New York, NY 10065, United States\\
$^{5}$School of Electrical and Computer Engineering, Cornell University and Cornell Tech, New York, NY 10044, United States
}

% ==================================== Abstract ====================================
\begin{abstract}
\noindent\textbf{Purpose:} 
To investigate whether an AI-based method can detect subtle inter-fraction changes in MR-Linac images acquired during radiotherapy and explore the broader potential of MR-Linac imaging.

\noindent\textbf{Methods:} 
This retrospective study included longitudinal 0.35T MR-Linac images from 761 patients. 
To identify temporal changes, we employed a deep learning model using temporal ordering via pairwise comparison, previously shown effective for longitudinal imaging studies.
The model was trained using first-to-last fraction pairs ($F_1$-$F_L$) and all pairs (\textit{All-pairs}). 
Performance was assessed using quantitative metrics (accuracy and AUC) and compared against a radiologist's performance.
Qualitative evaluation was performed using saliency maps, which identify anatomical regions associated with temporal imaging changes.

\noindent\textbf{Results:} 
The $F_1$–$F_L$ model demonstrated high performance (AUC=0.99; accuracy=0.95) and outperformed the radiologist in temporal ordering task.
The \textit{All-pairs} model also showed high performance (AUC = 0.97; accuracy = 0.91).
Regions contributing to predictions included the prostate, bladder, and pubic symphysis. 
The performance was correlated to fractional intervals and was reduced for non–radiation-exposed timepoints ($Sim$ and $F_1$), suggesting that observed changes may reflect both temporal variation and radiation exposure. 

\noindent\textbf{Conclusion:} 
MR-Linac imaging appears capable of capturing subtle changes during prostate radiotherapy that can be detected by AI models, even over approximately two-day intervals. The model’s high performance, together with quantitative and qualitative analyses, supports a potential role for MR-Linac in clinical applications beyond image guidance.

\end{abstract}
\vspace{-4.8mm}

\section*{Keywords}
Prostate Cancer, MR-Linac, Radiation Therapy, Longitudinal Image Analysis, Adaptive Radiotherapy, Deep Learning, Artificial Intelligence 

\section*{Highlights}
\begin{itemize}
    \item An AI model accurately predicted the temporal order of MR-Linac images during prostate radiotherapy.
    \item Performance differences between radiation-exposed and non–radiation-exposed timepoints suggest sensitivity to radiation-associated changes.
    \item Detected changes were most prominent in the prostate, with contributions from the bladder and pubic symphysis.
    \item Results suggest a potential role for MR-Linac imaging in treatment monitoring beyond image guidance.
\end{itemize}

% ==================================== Body ====================================

\section{Introduction}
\label{sec:intro}

A common therapeutic strategy for prostate cancer patients is external beam radiotherapy (EBRT), delivered either independently or in combination with surgery or hormone therapy \cite{mottet_eau-estro-siog_2017, mannerberg_dosimetric_2020}.
Image-guided radiotherapy has become standard practice for ensuring precise localization of the target and safe treatment delivery.
Magnetic Resonance-Linear Accelerator (MR-Linac) systems combine MRI with a linear accelerator, enabling the acquisition of high–soft-tissue–contrast MR images at every treatment fraction \cite{randall_towards_2022, ng_mri-linac_2023}. 
This integration allows visualization of the prostate, bladder, rectum, and adjacent tissues before radiation delivery and day-to-day anatomical changes throughout the treatment course \cite{murgic_mri-guided_2022,randall_towards_2022, ng_mri-linac_2023}.
Moreover, the MR images allow for adaptive therapy taking inter- and intra-fraction motion management into account, modifying each treatment fraction as needed based on spatiotemporal variations to ensure maximal target dose delivery with minimal dose to adjacent organs \cite{li_case_2022, sonke_adaptive_2019, randall_towards_2022, dona_lemus_adaptive_2024}. 

While MR-Linac provides unprecedented access to longitudinal MR imaging throughout the treatment course, potential treatment-induced changes in the images have been unexplored.
This is due to the relatively poor image quality and the subtle changes that are challenging for human observers to detect, combined with the fact that these images are primarily used for guiding radiotherapy and rarely reviewed beyond gross anatomical assessment. 
As a result, despite their potential clinical relevance for adaptive planning, treatment personalization, and toxicity prediction, systematic characterization of fraction-to-fraction changes remains limited \cite{dona_lemus_adaptive_2024}.

Several studies have reported significant prostate volume changes in diagnostic-quality MR images during a course of radiotherapy \cite{mannerberg_dosimetric_2020, alexander_prostate_2022}.
Longitudinal studies using multiparametric MRI \cite{algohary_longitudinal_2022, hanzlikova_longitudinal_2024} and quantitative MRI \cite{wang_longitudinal_2025} have demonstrated changes in prostate tumors and/or surrounding tissues before and after treatment delivery.
From MR-Linac images, prior works have shown changes in apparent diffusion coefficient (ADC) \cite{almansour_longitudinal_2023, fernando_repeatability_2024}. 
These prior studies suggest that MR imaging during radiotherapy can capture dynamic, patient-specific morphological and tissue changes.
However, to the best of our knowledge, inter-fraction changes in routinely acquired MR-Linac radiotherapy imaging have not been previously studied.

In this study, we developed an artificial intelligence (AI) model designed to detect subtle changes in MR images from MR-Linac fractions at short intervals (average of two days), adapting a longitudinal image comparison framework~\cite{kim_learning-based_2025}.
Following prior work, we hypothesized that a model capable of accurately ordering images chronologically would rely on features that evolve consistently with underlying clinical changes, which in this study correspond to cumulative radiation exposure.
The model achieved over $95\%$ accuracy, substantially outperforming an expert, and identified prostate and surrounding anatomical structures, such as bladder and pubic symphysis, as mainly altered regions during radiotherapy. 
These findings are supported by input ablation experiments and quantitative image analysis.
Overall, our study suggests that, beyond its established role in patient alignment and dose planning, MR-Linac imaging may have potential for studying longitudinal changes during radiotherapy, while AI-derived imaging features may help support treatment monitoring and adaptation.

\section{Methods}

\subsection{Data}
The study is HIPAA-compliant and IRB-approved with a waiver of informed consent for retrospective studies.
All data were de-identified prior to analysis. 

We retrospectively collected $1,071$ patients who underwent MR-guided radiotherapy for prostate cancer from 2018 to 2025, screened to 761 patients for the study (Details of selection are in the Section~\ref{sec:patientsel_detail} and flow diagram is in Figure~\ref{fig:split}).
The images include simulation scans (\textit{Sim} in Figure~\ref{fig:method}A) and corresponding treatment fractions ($F_1$ to $F_L$ in Figure~\ref{fig:method}A), with average inter-fraction intervals of two days (Figure~\ref{fig:hist_interdays}).
All images were acquired using a ViewRay MRIdian MR-Linac system (ViewRay Inc., USA) equipped with a 0.35-T MRI, using True Fast Imaging with Steady-State Free Precession (TRUFI) sequence.
The in-plane pixel spacing was $1.5 \times 1.5~\mathrm{mm}^2$, with a slice thickness of $1.5$ or $3~\mathrm{mm}$.
Apart from cropping centered on prostate masks (see Section~\ref{sec:seg_detail}
for segmentation method) to (80, 80, 80) sized volumes, no further pre-processing was performed.

Data were randomly split on a patient-wise basis into training (60\%, \textit{n} = 457), validation (20\%, \textit{n} = 152), and test data (20\%, \textit{n} = 152).
All MR images were cropped to dimensions of (80, 80, 80), centered at the centroid of the prostate mask (see Section~\ref{sec:seg_detail}
for segmentation method).

\subsection{Model}
The overall framework (Figure~\ref{fig:method}B) is based on Learning-based Inference of Longitudinal imAge Changes (LILAC) \cite{kim_learning-based_2025}, which learns meaningful changes from paired longitudinal images via predicting temporal order. 
As temporal ordering has demonstrated effectiveness in capturing monotonic changes, we adopted it to identify features consistently changing over the course of treatment. 
We employed a Siamese 3D convolutional neural network~\cite{bromley1994siamese, chopra_learning_2005, kim_learning_2023} with ResNet-18 \cite{He_2016_CVPR} architecture.
The model outputs logits for binary classification, predicting whether the input pair is in the correct or reversed temporal order.
These logits reflect the model's confidence, informed by learned differences between paired longitudinal images. 
Code is provided in \url{https://github.com/heejong-kim/LILAC}.
%\url{https://github.com/anonymized/mr-linac-ai/will_be_added}.
Further architectural details of the model are available in Section~\ref{sec:model_detail}.

For training, MR images from multiple fractions were paired in all possible ordered combinations, yielding $2,238$ pairs from 457 patients.
A curriculum learning was adopted, where initial training used only first-last fraction image pairs (\textit{$F_1$–$F_L$ pairs)}, which exhibit the largest radiation-induced differences and are therefore easier to learn~\cite{bengio_curriculum_2009}.
The model was further trained using all combinations of image pairs (\textit{All pairs}; \textit{All-pairs model}), with parameters initialized from those of the best $F_1$–$F_L$ model.
More details are in Section~\ref{sec:training_detail}.

\subsection{Assessment of Changes Across Fractions via Model Performance}
The held-out test data ($152$ patients, $732$ image pairs) were used to evaluate the performance across all experiments.

\paragraph{Temporal ordering performance evaluation.---}
Model performance was assessed using accuracy and the area under the receiver operating characteristic curve (AUC), with confidence intervals (CI) obtained from $1,000$ bootstrap resamples of the test data. 

\begin{itemize}
    \item \noindent\textbf{Comparison to expert:} 
A radiologist with over 15 years of experience performed a pairwise image comparison task to determine the correct temporal order of the same \textit{$F_1$–$F_L$} image pairs and assessed treatment changes.  
\item \noindent\textbf{Sim-F1 test:} 
We hypothesized that model performance on $Sim$ (not included in training) and $F_1$ pairs would be near random, on the premise that the model primarily captures treatment-related effects. 
As both images are acquired prior to the radiotherapy session for image guidance purposes, they are not exposed to radiation(Figure~\ref{fig:method}A,B).
\end{itemize}

\paragraph{Fraction-pairwise performance evaluation.---}
As shown in Kim et al.~\cite{kim_learning_2023}, the model output reflects the magnitude of changes between image pairs. 
Based on this observation, we examined logits of the \textit{All-pairs} model across fraction levels. 
To further quantify the relationship between fraction intervals and logits, we computed the Pearson correlation coefficient and conducted a linear mixed-effects model analysis to consider within-patient correlation (See Section~\ref{sec:lme_detail} for more details).

\subsection{Identification of Anatomical Regions with Changes Across Fractions Using AI Model}

\paragraph{Saliency map.---}
Saliency maps highlight regions that the model identifies as contributing most strongly to the differences between a given image pair, which are likely to reflect radiotherapy-induced effects. 
We used a modified Gradient-weighted Class Activation Mapping (Grad-CAM) \cite{selvaraju_grad-cam_2017, kim_learning-based_2025} to calculate the saliency maps. 
Implementation details and population-level visualization on an atlas are available in Section~\ref{sec:gradcam_detail}.

\paragraph{Input ablation.---}
Input ablation experiments were conducted to identify regions critical for the model’s prediction by examining model performance changes depending on input changes (Figure~\ref{fig:method}C). 
Occlusion was performed using either the segmented regions (see Section~\ref{sec:seg_detail}
for segmentation method) themselves (\textit{organ-masked-MR} input) or bounding boxes surrounding the segmented areas (\textit{box-masked-MR} input), where lower performance indicates stronger model reliance on the occluded organ regions.
Preservation was performed by retaining only the prostate, bladder, or both segmented regions in the MR images (\textit{only-organ-MR} input), where higher performance indicates stronger model reliance on the preserved organ regions.
These ablated inputs retain both organ size and intensity. 
To further assess the decoupled effect of organ size from intensity, segmented organ masks (prostate, bladder, or both) were used (\textit{mask} input).

\paragraph{Expert Saliency-Restricted MR Evaluation}
The radiologist performed the same temporal-order identification task using images restricted to the regions highlighted by the saliency maps, with the saliency maps themselves not shown (\textit{saliency-restricted-MR}; implementation details in Section~\ref{sec:saliency_restrict_detail}).
The radiologist also documented the rationale for each decision.
The overlaid saliency maps were later provided for expert's evaluation in order to interpret the model’s decision-making process. % and identify clinically relevant regions.

\subsection{Image Analysis}
Quantitative image analysis was performed to further investigate radiotherapy-induced changes and to compare them with the model and expert evaluations.
Changes in organ volume, and in the mean and standard deviation of image intensity for the prostate and bladder were assessed for all patients between $F_1$ and $F_L$. 
Prostate and bladder were defined using inferred masks generated by the deep learning model (See Section~\ref{sec:seg_detail} for details about segmentation).

\subsection{Statistical Analysis}
An independent two-sided t-test was conducted for comparisons of scores (accuracy and AUC) and patient characteristics.
The Wilcoxon signed-rank test was used to compare organ volume and intensity from $F_1$ to $F_L$.
For all statistical tests, p-value of 0.05 was used as the threshold for statistical significance.
See Section~\ref{sec:stat_detail} 
for details on the tools used.

\section{Results}

\subsection{Patient Characteristics}
All patients in the cohort (Table~\ref{tab:patient}) were diagnosed with elevated prostate-specific antigen (PSA) levels and prostate cancer.
All patients except 43 patients received five fractions of MR-Linac treatment, mostly administered every two days (Figure~\ref{fig:hist_interdays}).
Approximately 40\% of patients in the cohort underwent androgen deprivation therapy (ADT). 
There were no statistically significant differences between the train, validation, and test splits.

\subsection{Evaluation of Changes in Images across Fractions via Model Performance}
Because temporal order is inferred from image differences, output logits were used as a surrogate measure of change across radiotherapy fractions.

\paragraph{Model performance and comparative analysis.---} 
The model performance on the test set was quantitatively evaluated (Table~\ref{tab:performance}).
The $F_1$-$F_L$ model yielded the accuracy of 0.95 (95\% CI: 0.93, 0.98), and the AUC of 0.99 (95\% CI: 0.99, 1.00).
The model outperformed the expert achieving an accuracy of 0.82 (95\% CI: 0.76, 0.88), of which the difference was statistically significant (independent two-sided t-test, $p < .0001$).

The \textit{All-pairs} model achieved an accuracy of 0.95 (95\% CI: 0.93, 0.97) and an AUC of 0.99 (95\% CI: 0.99, 1.00).
In contrast, inference on $Sim$-$F_1$ pairs yielded a significantly lower accuracy of 0.40 (95\% CI: 0.34, 0.45) and AUC of 0.34 (95\% CI: 0.28, 0.40) (independent two-sided t-test, $p < .0001$).
The model achieved high accuracy on images acquired during radiotherapy but performed poorly on pre-radiotherapy images ($Sim$-$F_1$), indicating that radiotherapy-associated changes may underlie the model’s decisions.
Patients who underwent ADT showed no statistically significant differences in model logits (independent two-sided t-test), further supporting that the detected temporal changes are potentially associated with radiotherapy.

\begin{figure*}[p]
  \centering
  \includegraphics[width=\textwidth]{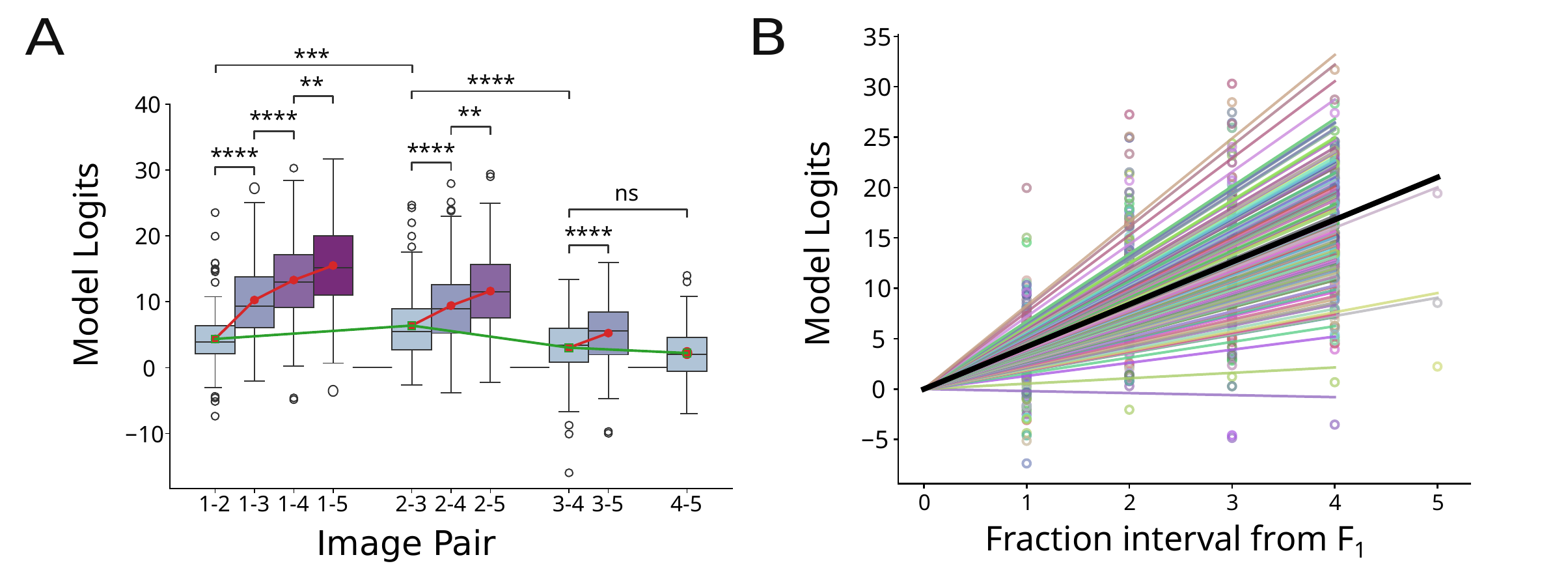}
  \caption{\textbf{Pairwise performance} (A) Model logits of the \textit{All-pairs} model for different image pairs. Model logits reflect the model's confidence in its predictions, which can be interpreted as the magnitude of changes detected between paired images. Pairs of numbers on the x-axis indicate the fractions at which the paired images were acquired. Cases were grouped by the first fraction and ordered and color-coded by the interval between the paired fractions. Stars indicate statistical significance between compared cases (independent two-sided t-test; ns: $.05 < p$, **: $.001 < p \le .01$, ***: $.0001 < p \le .001$, ****: $p \le .0001$). Means within the same first fraction group are shown in red, whereas means with the same fraction interval but different first fraction groups are shown in green.
  (B) Linear mixed-effects model showing the relationship between fraction intervals from the first fraction and model logits (black), with patient-wise gradients shown in multicolor.}
  \label{fig:score_pair}
\end{figure*}

\paragraph{Association between model performance and radiotherapy course.---} 
We hypothesize that image differences between radiotherapy fractions would increase with longer fraction intervals due to potential cumulative radiation exposure, if the detected temporal changes are associated with radiotherapy.
To examine this, the performance of the \textit{All-pairs} model for each fraction pair was evaluated (Figure~\ref{fig:score_pair} and Figure~\ref{fig:figure_score_pair_AB}).
Model logits were positively correlated with the inter-fraction interval (Pearson $r=0.59$, $p < .0001$), as well as accuracy and AUC all increased proportionally with the fraction intervals (Figure~\ref{fig:figure_score_pair_AB}).
Within groups of image pairs sharing the same first image (e.g. 1-2, 1-3, 1-4, and 1-5), the mean model logits increased with the fraction interval (red lines in Figure~\ref{fig:score_pair}A).
Among adjacent image pairs (1-2, 2-3, 3-4, and 4-5), the $F_2$-$F_3$ pairs exhibited significantly higher than other one-interval pairs (green lines in Figure~\ref{fig:score_pair}A), 
suggesting that changes are most pronounced between the second and third fractions.
These results demonstrate that the model predictions can quantify changes associated with radiotherapy between fractions.

Incorporating individual variability in the linear mixed-effects model resulted in a significant difference from the reduced model without random effects, according to the likelihood ratio test ($p < .0001$) (Figure~\ref{fig:score_pair}B).
This indicates that the changes vary across patients, suggesting heterogeneous treatment effects.

\subsection{Localization of Changes Driving Model Predictions }\label{sec:region}
\paragraph{Saliency Map.---} 
Modified Grad-CAM\cite{selvaraju_grad-cam_2017,kim_learning_2023} based saliency maps were used to interpret which regions contribute to the model's predictions.
According to the radiologist's evaluation, the saliency peaks most frequently occurred in the bladder lumen (20.53\%) and pubic symphysis (19.21\%), with the prostate commonly involved (Table~\ref{tab:expert_heatmap}). 
Group-level visualization of the heatmaps on the atlas of all patients is shown in Figure~\ref{fig:gradcam} (Peak distribution of the heatmaps in Figure~\ref{fig:atlas_gradcam_peak}),
of which the primary peak regions are the pubic symphysis and bladder, while the area spans multiple anatomical regions, including the prostate.

\paragraph{Input Ablation.---} 
Input ablation experiments were conducted to identify anatomical regions critical to the model’s performance (Table~\ref{tab:ablation}).
All input ablation test results were significantly different from the baseline result. 
Occluding both prostate and bladder regions in MR (\textit{Organ-masked-MR} and \textit{Box-masked-MR}) shows a significant drop in the performance, indicating that prostate, bladder, and surrounding box regions contain important radiotherapy-related changes. 
\textit{Box-masked-MR} caused greater performance degradation than \textit{Organ-masked-MR}, possibly suggesting the significance of shape changes in model prediction, which motivated the evaluation of \textit{Mask} input.

Preservation experiments show the effect of organ shape and texture as input (\textit{Only-organ-MR}) and only shape as input (\textit{Mask}). 
Using masks alone (\textit{Mask}) reduced performance relative to \textit{Only-organ-MR}, although accuracy remained above 0.9 ($F_1$–$F_L$ model) and 0.8 (\textit{All-pairs} model). 
Prostate mask had a larger impact than bladder mask.
The results suggest that prostate size is a major contributor, but intensity also contributes.
The performance degradation relative to the baseline was greater for \textit{All-pairs} model than $F_1$–$F_L$ model, indicating that shape and texture information are essential to capture subtle inter-fraction changes.

\textit{All-pairs} model performed best when both prostate and bladder were preserved, whereas $F_1$–$F_L$ model achieved higher performance when only prostate was preserved.
This indicates that the most pronounced changes between pre- and post-treatment are localized to the prostate, whereas more subtle inter-fraction changes involve both prostate and bladder.

\paragraph{Expert Saliency-Restricted-MR Evaluation.---}
The expert additionally assessed the model’s saliency maps through performing a prediction test on \textit{saliency-restricted MR} images (Prediction Test on Saliency-Restricted-MR in Figure~\ref{fig:method}C).
Although the FOV of the images is limited, the expert achieved an accuracy of 0.72 (95\% CI: 0.66–0.79), showing relevance of model-highlighted regions for capturing changes.

For full volume prediction test, the expert specified the decision criteria in descending order of importance: (1) darkening of the prostate, (2) reduced distinction between the peripheral zone and transition zone borders, (3) enlargement of the prostate, and (4) reduced definition of the prostate’s outer edge.
The expert primarily relied on changes in prostate volume and MR image characteristics, including prostate brightness and clarity of the prostate edges.
The rationales were similar for \textit{saliency-restricted-MR} test (Table~\ref{tab:expert_reason}).

\subsection{Organ Volume and Intensity Changes in Images During Radiotherapy}
Changes in prostate and bladder volume and intensity between $F_1$ and $F_L$ were analyzed to validate the model and expert findings (Figure~\ref{fig:organ}).
Prostate volume increased significantly from $F_1$ to $F_L$, whereas bladder volume decreased significantly (one-sided Welch's t-test, $p < .0001$).
Both organs became darker, with the mean intensity in each region decreasing significantly from $F_1$ to $F_L$ (independent one-sided t-test, $p = .02$ and $.001$, respectively).
Additionally, the variance of intensity, reflecting heterogeneity, increased significantly in the prostate, whereas it decreased significantly in the bladder (one-sided independent t-test, $p = .002$ and $p < .0001$, respectively).

The observed increase in prostate volume and reduction in brightness aligns with both the model’s results and the expert’s assessments.
These results suggest that the model’s detection of inter-fractional changes may be partially explained by changes in prostate volume and intensity.

\section{Discussion}
In this study, we demonstrate that AI-based analysis of prostate MR-Linac images may enable detection of treatment-related changes during radiotherapy. This study characterizes changes in MR-Linac images across fractions over short time scales--one or a few days between consecutive fractions. Analyzing changes in MR-Linac images is challenging and has received limited attention, as many inter-fraction changes are subtle and difficult to detect visually. We not only developed an AI model that accurately predicts the temporal order of MR images acquired during MR--Linac treatment fractions separated by one to several days, but also conducted extensive evaluations--including model input ablation and clinician assessments--to validate the model-detected changes.

The model achieved near‑perfect performance, outperforming expert assessment (Table~\ref{tab:performance}), suggesting the presence of subtle differences that are difficult to detect visually. 
Its performance on image pairs not exposed to radiotherapy (\textit{$Sim$–$F_1$}) was significantly lower than on radiotherapy-affected pairs, suggesting a possible association between the detected changes and radiation exposure.
Performance also increased with longer inter‑fraction intervals, consistent with cumulative radiation effects (Figure~\ref{fig:score_pair}). 
No significant performance differences were observed between patients with and without ADT, further suggesting a primary role for radiotherapy in the observed temporal changes.
Failure cases occurred mainly in consecutive‑fraction pairs and were not associated with patient characteristics or imaging parameters.

\begin{figure}[p]
  \centering
  \includegraphics[width=0.8\textwidth]{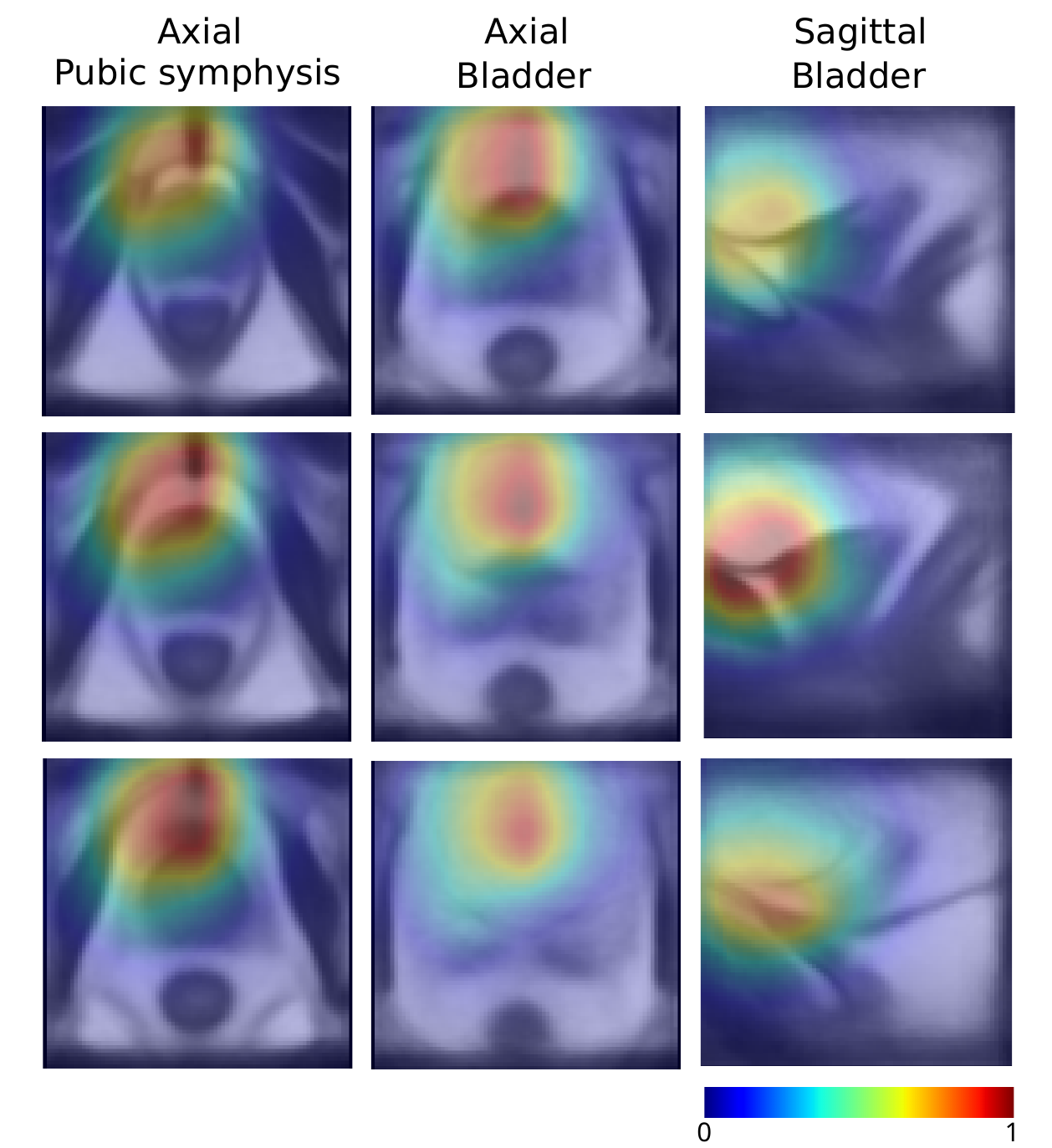}
  \caption{\textbf{Saliency map on an atlas.} The atlas was built from $F_1$ of all patients in the test data. The visualized heatmap is the mean of saliency maps for all patients in the test data, obtained from \textit{All-pairs} model inference on $F_1$-$F_L$ pairs and transformed into the atlas space. Column titles indicate the slice orientations (axial, sagittal) and the primary regions of interest highlighted by the saliency maps, although these regions span multiple organs.}
  \label{fig:gradcam}
\end{figure}

Anatomical regions detected by the model were investigated through saliency maps and input ablation.
The saliency maps mainly highlighted the pubic symphysis, bladder, and prostate (Figure~\ref{fig:gradcam}, Table~\ref{tab:expert_heatmap}).
This is consistent with reports that some patients who had radiotherapy or MR-Linac treatment for prostate cancer experience genitourinary symptoms such as urinary frequency, urethral stricture, and cystitis \cite{liu_clinical_2023, pisani_urinary_2022, willigenburg_accumulated_2022}.
While less has been reported on the pubic symphysis, pubic bone osteomyelitis and changes in T2- and T1-weighted signals of the involved pubic rami were reported in patients who had radiotherapy for prostate cancer \cite{sexton_magnetic_2019}.
These results suggest that the AI model may capture radiotherapy-affected regions, including potential side effects, and that the detected changes may be associated with radiation exposure.

Quantitative analysis confirmed that the prostate enlarged and darkened with increased intensity heterogeneity, while the bladder shrank, darkened, and intensity became more uniform (Figure~\ref{fig:organ}).
Radiotherapy may have contributed to prostate enlargement and associated bladder compression, possibly reflecting treatment-related inflammatory changes. 
Signal darkening occurred in both, but increased intensity variability was observed only in the prostate, suggesting different responses from targeted versus non-targeted radiation.

Several limitations should be acknowledged. Grad-CAM may yield broad activation patterns; fine-grained explainability methods could improve localization. This proof-of-concept used data from a single platform, and performance on other platforms remains to be determined. 
Our findings suggest a possible association between the detected temporal imaging changes and radiotherapy. However, further investigation is needed to determine whether these reflect radiation-specific biological effects.

In conclusion, we show that AI can reliably detect temporal changes in prostate MR‑Linac images, and that these changes may reflect effects of radiotherapy.
These results support the feasibility of AI‑based identification of in‑treatment imaging changes, which may serve as potential imaging biomarker of treatment response or adverse effects. 
AI‑derived imaging features may support future applications in treatment monitoring, adaptation, and personalized radiotherapy. 
Future work will evaluate whether the model can predict biochemical failure and adverse effects, with the goal of improving patient selection for primary and adjuvant therapies.

\section{Acknowledgement}
This work was supported by the National Institutes of Health [grant numbers xxxx, yyyy, and zzzz - anonymized for the review].
%NIH grants K25CA283145, R01AG053949, R01AG064027, and R01AG070988.

{
    \small
    \bibliographystyle{vancouver}
    \bibliography{main}
}

% ==================================== Table & Figure ====================================

\clearpage
\begin{table}[p]
\centering
\begin{tabularx}{\linewidth}{Xccc}
\toprule
Characteristic & Training Data & Validation Data & Test Data \\
\midrule
No. of patients & 457 & 152 & 152 \\
Age (y) & 73 (10) & 73 (11) & 72.5 (10.25)   \\
No. of fractions per patient & 5 (0) & 5 (0) & 5 (0) \\
Days between fractions & 2 (1) & 2 (1) & 2 (1) \\
PSA level (ng/mL) & 2.36 (4.38) & 2.13 (4.02) & 2.21 (2.59) \\
No. of patients with ADT & 180 (39) & 66 (43) & 55 (36) \\
\bottomrule
\end{tabularx}
\caption{\textbf{Patient Characteristics.} Age, the number of fractions per patient, days between fractions, and PSA level are reported as medians with interquartile ranges in parentheses.
The number of patients with ADT is reported as medians with percentages in parentheses. No significant differences were found among values from the training, validation, and test data, except for the number of patients (independent two-sided t-test).}
\label{tab:patient}
\end{table}

\clearpage
\begin{table}[p]
\centering
\begin{tabularx}{\linewidth}{l >{\raggedright\arraybackslash}X c c}
\toprule
Pair & Performer & ACC & AUC \\
\midrule
$F_1$-$F_L$ & AI & 0.95 (0.93, 0.98) & 0.99 (0.99, 1.00) \\
 & Expert & 0.82 (0.75, 0.87)$^*$ & - \\
\midrule
\textit{All-pairs} & AI  & 0.91 (0.90, 0.92) & 0.97 (0.96, 0.97) \\
& AI, inference for $Sim$-$F_1$ pairs  & 0.39 (0.34, 0.45)$^*$ & 0.34 (0.28, 0.40)$^*$ \\
\bottomrule
\end{tabularx}
\caption{\textbf{Performance of the model and the expert.} Star ($^*$) shows statistically significant differences compared to previous rows (independent two-sided t-test, $p < .0001$). 
High performance (accuracy and AUC $>$ 0.90) suggests that significant imaging changes exist across prostate radiotherapy using MR-Linac. 
Notably, the significant performance degradation when inferring on \textit{$Sim$-$F_1$} pairs (both images acquired before radiotherapy) indicates that the detected temporal changes are associated with the progression of the radiotherapy course.}
\vspace{-0.5cm}
\label{tab:performance}
\end{table}

\clearpage
\begin{table*}[p]
\centering
\begin{tabularx}{\textwidth}{X p{3.2cm} c c c c} %{l X c c c c}
\toprule
 & & \multicolumn{2}{c}{$F_1$-$F_L$ model} & \multicolumn{2}{c}{All-pairs model} \\
\cmidrule(lr){3-4} \cmidrule(lr){5-6}
Input & Organ & ACC & AUC & ACC & AUC \\
\midrule

Baseline & - & \textbf{0.95 $\pm$ 0.01}$^*$ & \textbf{0.99 $\pm$ 0.00}$^*$ & \textbf{0.91 $\pm$ 0.01}$^*$ & \textbf{0.97 $\pm$ 0.00}$^*$  \\

\midrule
Organ- & Prostate & -0.03 $\pm$ 0.01 & -0.01 $\pm$ 0.00 & -0.01 $\pm$ 0.01 & -0.00 $\pm$ 0.00 \\
masked- & Bladder & -0.02 $\pm$ 0.01 & -0.01 $\pm$ 0.00 & -0.04 $\pm$ 0.01 & -0.02 $\pm$ 0.00 \\
MR$\hspace{0.3cm}(\downarrow)$ & Prostate $\cup$ Bladder & \textbf{-0.05 $\pm$ 0.02} & \textbf{-0.04 $\pm$ 0.01} & \textbf{-0.04 $\pm$ 0.01} & \textbf{-0.03 $\pm$ 0.00} \\

\midrule
Box- & Prostate & -0.07 $\pm$ 0.02 & -0.06 $\pm$ 0.02 & -0.04 $\pm$ 0.01 & -0.02 $\pm$ 0.00 \\
masked- & Bladder & -0.04 $\pm$ 0.02 & -0.02 $\pm$ 0.01 & -0.08 $\pm$ 0.01 & -0.06 $\pm$ 0.01 \\
MR$\hspace{0.3cm}(\downarrow)$ & Prostate $\cup$ Bladder & \textbf{-0.20 $\pm$ 0.02} & \textbf{-0.19 $\pm$ 0.02} & \textbf{-0.25 $\pm$ 0.01} & \textbf{-0.25 $\pm$ 0.01} \\

\midrule
Only- & Prostate & \textbf{-0.03 $\pm$ 0.02} & \textbf{-0.01 $\pm$ 0.01 }& -0.10 $\pm$ 0.01 & -0.08 $\pm$ 0.01 \\
organ- & Bladder & -0.09 $\pm$ 0.02 & -0.06 $\pm$ 0.01 & -0.14 $\pm$ 0.01 & -0.12 $\pm$ 0.01 \\
MR$\hspace{0.3cm}(\uparrow)$ & Prostate $\cup$ Bladder & -0.03 $\pm$ 0.02 & -0.02 $\pm$ 0.01 & \textbf{-0.08 $\pm$ 0.01} & \textbf{-0.05 $\pm$ 0.00} \\

\midrule
Mask & Prostate & \textbf{-0.05 $\pm$ 0.02} & \textbf{-0.04 $\pm$ 0.01} & -0.16 $\pm$ 0.01 & -0.15 $\pm$ 0.01 \\
($\uparrow$) & Bladder & -0.21 $\pm$ 0.03 & -0.22 $\pm$ 0.03 & -0.29 $\pm$ 0.01 & -0.30 $\pm$ 0.01 \\
& Prostate $\cup$ Bladder & -0.09 $\pm$ 0.02 & -0.07 $\pm$ 0.02 & \textbf{-0.12 $\pm$ 0.01} & \textbf{-0.13 $\pm$ 0.01} \\

\bottomrule
\end{tabularx}
\caption{\textbf{Input ablation.} Values in the first row are presented as mean $\pm$ standard deviation. 
All other values represent the difference from the baseline mean $\pm$ standard deviation.
$^*$: All cases are significantly different from the baseline results (independent two-sided t-test, $p < .0001$). 
$\downarrow$: Performance degradation compared to the baseline indicates strong model dependence on the targeted organ regions. The worst case is highlighted in bold.
$\uparrow$: Higher performance indicates strong model dependence on the targeted organ regions. The best case is highlighted in bold.}
\vspace{-0.5cm}
\label{tab:ablation}
\end{table*}

\clearpage
\begin{figure}[p]
  \centering
  \includegraphics[width=0.85\linewidth]{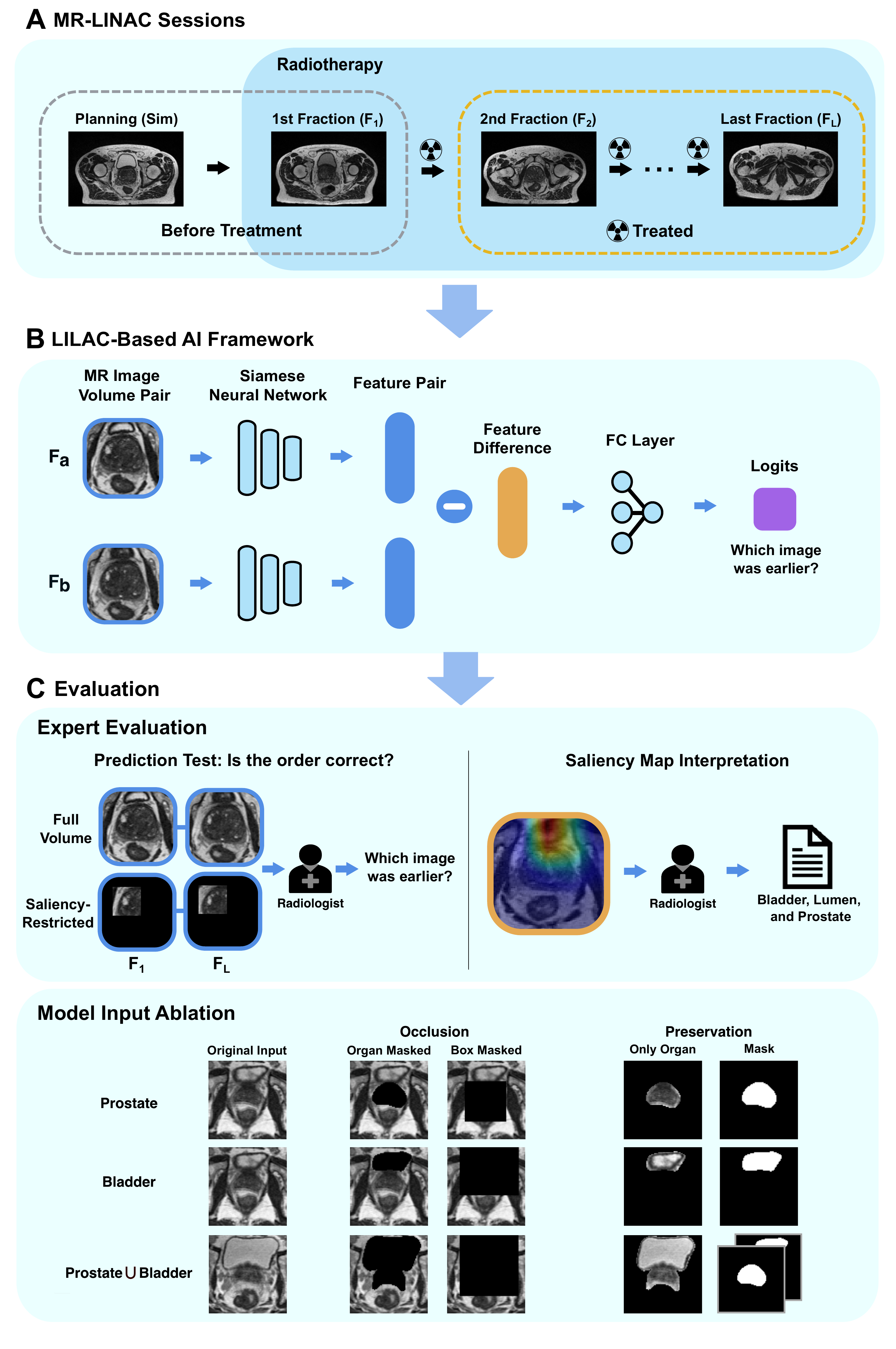}
  \caption{\textbf{Overview of the MR-Linac workflow and AI framework.}}
\label{fig:method}
\end{figure}

\begin{figure*}\ContinuedFloat
  \caption{(Continued) (A) Process of MR-Linac sessions across multiple fractions. \textit{Sim} refers to the simulation scans obtained prior to MR-Linac treatment sessions. $F_1$, $F_2$, ..., $F_L$ represent MR images acquired at the first, second, ..., and last fractions, respectively, prior to image-guided radiotherapy. 
  (B) Schematic diagram of the LILAC-based AI model. Paired 3D MR image volumes from fractions are input to a Siamese 3D convolutional neural network with ResNet-18 \cite{He_2016_CVPR} architecture, in which features of each image are obtained. The differences between these two features are fed into a fully connected layer. The logits are used for binary classification to predict if the image pairs are in the correct temporal order or not. 
  (C) Experimental design for evaluation. A radiologist performed the same temporal-ordering task first with full volumes and next with saliency-restricted images. The radiologist also evaluated anatomical regions highlighted by the saliency maps. 
  Model input ablation experiments were designed to evaluate altered anatomical regions. Performance changes when the prostate, bladder, or both regions or the surrounding box regions were masked in MR images (Organ Masked, Box Masked) were evaluated, as well as performance for partial MR images encompassing the organ regions (Only Organ). Performance for using only masks (Mask) was evaluated to investigate the effects of the organ shapes, compared to MR intensities.}
  %\label{fig:method}
\end{figure*}

\clearpage

\clearpage

% ==================================== Supplementary ====================================
\clearpage
%\appendix
\setcounter{section}{0}
\renewcommand{\thesection}{S\arabic{section}} % Prefix sections with "A"
\setcounter{figure}{0}
\renewcommand{\thefigure}{S\arabic{figure}}   % Prefix figures with "A"
\setcounter{table}{0}
\renewcommand{\thetable}{S\arabic{table}}     % Prefix tables with "A"

\begin{center}
    \Large \textbf{Supplemental Material}
\end{center}

\section{Patient screening}
\label{sec:patientsel_detail} 
\begin{figure*}[p]
\begin{center}
\includegraphics[width=0.7\linewidth]{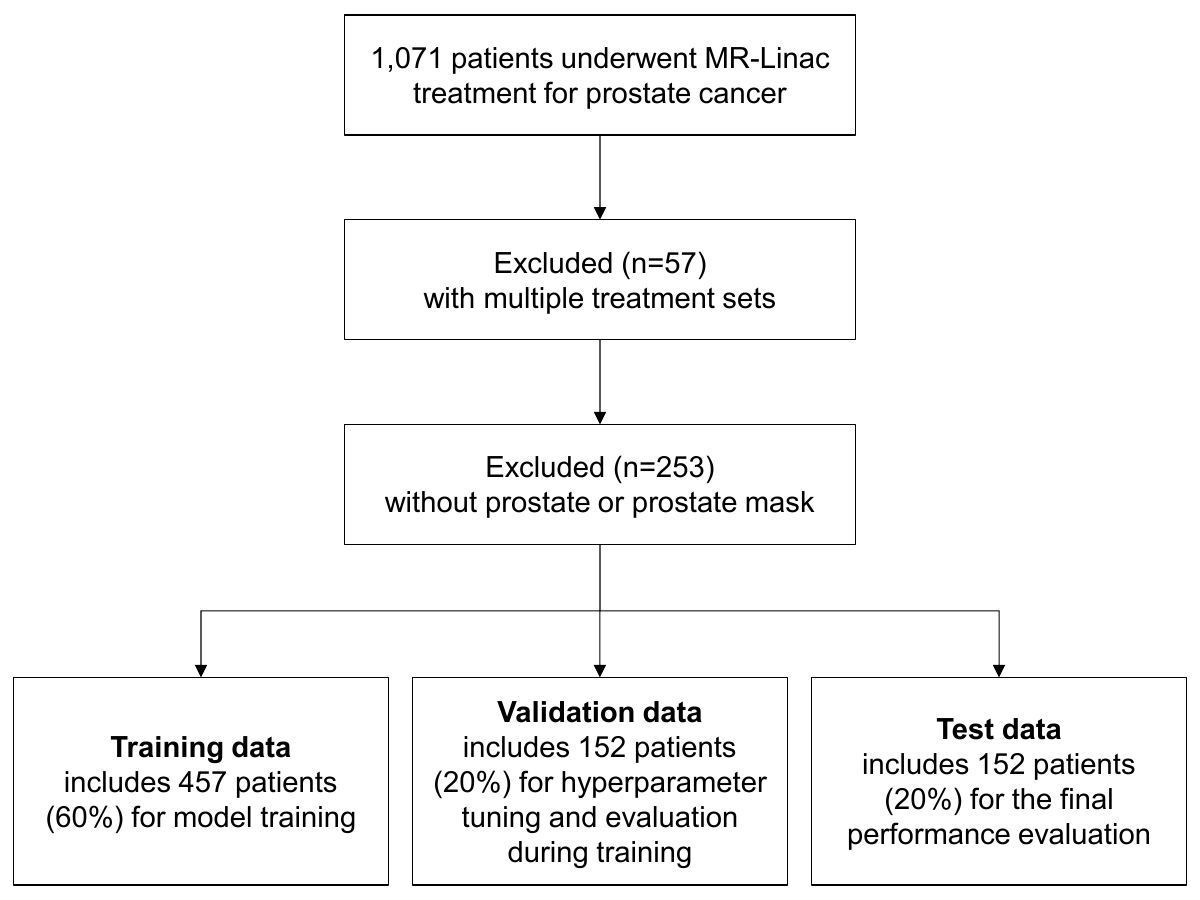}
\end{center} 
\caption{\textbf{Flow diagram of cases, indicating inclusion and exclusion of patients, and random split of the data. }}
\label{fig:split}
\end{figure*}

Of the 1,071 patients, those who underwent multiple treatment sets (\textit{n} = 57) and those without a prostate or prostate mask (\textit{n} = 253) were excluded (Figure~\ref{fig:split}), as their imaging characteristics may differ from the rest of the cohort. 

\section{Segmentation of Organs in MR Images from MR-Linac Fractions}
\label{sec:seg_detail} 
The prostate, bladder, and rectum are routinely contoured in radiotherapy guide images obtained before treatment by radiation oncologists for treatment planning purposes. 
Since contouring is routinely performed only on the guide images, we trained an nnU-Net~\cite{isensee2021nnu} on the guide images to get prostate, bladder, and rectum masks in MR images from MR-Linac fractions. 
We assumed that the model would generalize to images from fractions, as both were obtained from the same MR-Linac system and target the same anatomical regions with consistent imaging protocols. 

The segmentation model was trained, validated, and tested using split datasets of guide MR images with corresponding prostate, bladder, and rectum masks delineated by radiation oncologists.
On the test set, the Dice scores of the model were 0.85, 0.95, and 0.82 for prostate, bladder, and rectum.
The trained model was then applied to MR images acquired during MR-Linac treatment fractions to generate organ masks.
Partial qualitative evaluation of the resulting masks, together with quantitative comparison between the trained MR-Linac temporal-ordering models, demonstrated that this approach outperformed masks obtained via image registration.

We used dilated masks for input ablation experiments, except for the mask-only condition, to ensure that organ boundaries were included in the occluded or preserved regions.
Binary masks were morphologically dilated by two voxels in all directions using a 3D 26-connected structuring element.

\section{Model Architecture Details}
\label{sec:model_detail} 
The overall framework (Figure~\ref{fig:method}B) is based on learning-based inference of longitudinal image changes (LILAC) \cite{kim_learning-based_2025}, which extracts biologically meaningful temporal changes directly from paired longitudinal images. 
It has shown effectiveness across multiple longitudinal imaging applications, including embryo development, wound healing, brain aging, and Alzheimer’s disease.
Adapting the LILAC framework, we employed a convolutional neural network (ResNet-18 \cite{He_2016_CVPR}) with a Siamese architecture \cite{bromley1994siamese, chopra_learning_2005, kim_learning_2023}, which takes a pair of images as input and extracts the corresponding feature representations to predict the correct order of the given longitudinal image pair.
For Siamese architecture, ResNet-18 \cite{He_2016_CVPR} with 3D convolutional layers was used, which demonstrated superior performance compared to simple convolutional neural networks (Table~\ref{tab:performance_sup}). 
The difference between these feature vectors is subsequently fed to a bias-free fully connected layer, enabling the model to learn differences between the two images.
The resulting logits are used for binary classification to determine whether the input pair is in the correct or reversed temporal order using the binary cross-entropy loss function.

\section{Training Details}
\label{sec:training_detail} 
For the temporal-ordering task on image pairs, MR images from fractions were paired in all possible permutations with repetition.
Consequently, $n^2$ image pairs were generated per patient ($n$: the number of fractions for the patient). 
In total, this pairing yielded $2,238$ pairs for the training data, $737$ pairs for the validation data, and $732$ pairs for the test data. 
Ground truth labels were assigned as 1 for the temporally ordered, 0 for the temporally reversed, and 0.5 for the identical pairs.

A curriculum learning approach was adopted as part of the training strategy~\cite{bengio_curriculum_2009}, where a model is gradually exposed to examples in an order from easy to hard to improve learning efficiency and performance.
The approach resulted in improved performance compared with training from scratch (Table \ref{tab:performance_sup}).

Models were trained for 100 epochs, with the optimal epoch identified as the one yielding the smallest validation loss.
Training and validation losses were monitored to check for convergence and overfitting.
Model parameters were optimized using the Adam optimizer \cite{kingma_adam_2014} to minimize the binary cross-entropy loss between the model outputs and the ground truth.
The learning rate was initially set to 0.001 and decayed by a factor of 0.5 every 20 epochs.
Training was conducted using the \textit{Python PyTorch} library on a NVIDIA A40 GPU.

\section{Linear Mixed-Effects model}
\label{sec:lme_detail} 
A linear mixed-effects model was constructed to relate the positive fraction intervals from the first fraction in $F_1$-$F_X$ ($X>1$) pairs and the corresponding model logits. 
Because there was no change when comparing the initial time point to itself, intercepts were omitted from the linear mixed-effects model, and only subject-specific random slopes were included. 
The significance of individual variability was evaluated using a likelihood ratio test comparing the full model with random effects to a reduced model without them.

\section{Grad-CAM}
\label{sec:gradcam_detail}
A modified Gradient-weighted Class Activation Mapping (Grad-CAM) \cite{selvaraju_grad-cam_2017, kim_learning-based_2025} was calculated to obtain the saliency maps, which visualize the areas in inputs on which the model largely depends for prediction. 
Gradients of the model outputs were computed with respect to the final activation maps, and importance was assigned at each spatial location individually. 
The activation maps were then weighted element-wise by the corresponding gradients to produce the saliency maps.
Regions with high Grad-CAM values were analyzed to identify areas likely affected by MR-Linac radiotherapy.

For group-level visualization of the saliency maps, $F_1$ atlas was constructed from all patients in the test data using \textit{ants.registration.build\_template} module in \textit{ANTsPy} \cite{tustison_antsx_2021}, python implementation of Advanced Normalization Tools (ANTs). 
The module begins with an initial template, which is the average of all images.
All images were nonlinearly registered using SyN (Symmetric Normalization)---ANTs’ diffeomorphic registration model---without rigid or affine alignment, over three iterations. 
At each iteration, warped images were averaged and blended with the previous template using a blending weight of 0.75 to ensure stable convergence, with deformation updates controlled by a gradient step of 0.2, yielding an unbiased diffeomorphic mean representation of the cohort.
Using the constructed atlas, patient-wise saliency maps were transformed into the atlas space, averaged across images, and overlaid for visualization (Figure~\ref{fig:gradcam}). 

\section{Saliency-Restricted-MR Details}
\label{sec:saliency_restrict_detail}
Binary masks were generated by thresholding the min-max scaled saliency maps at 0.3.
Using these masks, we created bounding boxes that encompassed the thresholded regions.
The resulting cropped images were provided to experts for the \textit{saliency-restricted-MR} evaluation.

\section{Statistical Analysis Details}
\label{sec:stat_detail}
An independent two-sided t-test (\textit{Python scipy.stats.ttest\_ind}) was conducted for pairwise comparisons of scores (accuracy and AUC) and patient characteristics.
The Wilcoxon signed-rank test (\textit{Python scipy.stats.wilcoxon}) was used to compare organ volume and intensity from $F_1$ to $F_L$.

\clearpage
\begin{table*}[p]
\centering
\begin{tabularx}{\textwidth}{c X r} 
\toprule
\textbf{Index} & \textbf{Expert Evaluation of Grad-CAM Peak} & \textbf{\# of Cases} \\
\midrule
1 & Bladder lumen & 31 (20.53\%) \\
2 & Pubic symphysis & 29 (19.21\%) \\
3 & Bladder lumen, public symphysis, and periprostatic fascia (1 + 2 + 8) & 18 (11.92\%) \\
4 & Pubic symphysis and bladder lumen (1 + 2) & 17 (11.26\%) \\
5 & Base of prostate at urethral inlet & 8 (5.30\%) \\
6 & Public symphysis and base of prostate at urethral inet (2 + 5) & 8 (5.30\%) \\
7 & Pubic symphysis and periprostatic fascia (2 + 8) & 7 (4.64\%) \\
8 & Periprostatic fascia & 6 (3.97\%) \\
9 & Bladder wall & 5 (3.31\%) \\
10 & Bladder lumen and lateral prostate (1 + 14) & 4 (2.65\%) \\
11 & Random fascia above the bladder & 4 (2.65\%) \\
12 & Urethral inlet and periprostatic fascia (5 + 8) & 3 (1.99\%) \\
13 & Bladder apex & 2 (1.32\%) \\
\bottomrule
\end{tabularx}
\caption{\textbf{Expert evaluation of peak regions of saliency maps from the \textit{All-pairs} model on $F_1$-$F_L$ pairs.} Regions with only one case are omitted from the table: prostate peripheral zone, pubic symphisis and lateral prostate possibly peripheral zone (2 + 14),  anterior prostate and possibly artifact, pubic symphysis and tissue above bladder (2 + other), urethral inlet and bladder lumen (1 + 5), like 3 but ill-defined, ill-defined areas spanning the pubic symphysis and prostate base with additional bilateral areas in fascia above the bladder, periprostatic fascia and fascia above the bladder (8 + 11), and bladder lumen and periprostatic fascia (1 + 8).}
\label{tab:expert_heatmap}
\end{table*}

\clearpage

\begin{table*}[p]
\centering
\begin{tabularx}{\linewidth}{X r}
\toprule
\textbf{Expert Rationale} & \textbf{\# of Cases} \\
\midrule
Areas in the PZ get darker. & 85 (56\%) \\
Areas in the TZ get darker. & 51 (34\%)\\
TZ features get indistinct. & 13 (9\%) \\
Prostate missing but MR signals get darker& 2 (1\%) \\
\bottomrule
\end{tabularx}
\caption{\textbf{Expert rationales for temporal ordering decisions on \textit{saliency-restricted-MR} images.}}
\label{tab:expert_reason}
\end{table*}

\clearpage
\begin{table*}[p]
\centering
\begin{tabularx}{\linewidth}{l >{\raggedright\arraybackslash}X c c}
\toprule
Pair & Model \& Training Scheme & ACC & AUC \\
\midrule
$F_1$-$F_L$ & CNN-3D & 0.89 & 0.96 \\
\midrule
\textit{All} & CNN-3D, Curriculum learning from the $F_1$-$F_L$ model & 0.89 & 0.96 \\
 & CNN-3D, From scratch & 0.88 & 0.95 \\
 & ResNet-18-3D, From scratch & 0.88 & 0.95 \\
\bottomrule
\end{tabularx}
\caption{\textbf{Performance of various models and training schemes.} CNN-3D consists of four blocks of 3D convolutional layer (number of output channels set as 16, 32, 64, 16, respectively), batch normalization, leaky ReLU, and average pooling. The ResNet-18-3D model trained with curriculum learning, which was used in the study for subsequent analyses, demonstrated superior performance compared with the CNN-3D and the ResNet-18-3D model trained from scratch.}
\vspace{-0.5cm}
\label{tab:performance_sup}
\end{table*}

\clearpage

\clearpage

\begin{figure*}[t]
\begin{center}
\includegraphics[width=\linewidth]{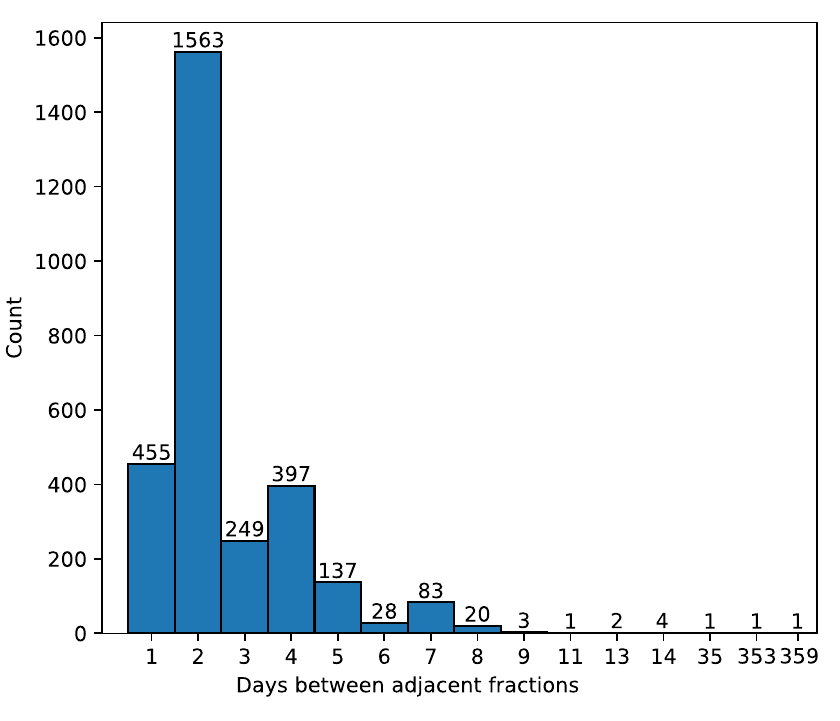}
\end{center} 
\caption{\textbf{Histogram of days between adjacent fractions.}}
\label{fig:hist_interdays}
\end{figure*}

\clearpage

\begin{figure*}[p]
\begin{center}
\includegraphics[width=\linewidth]{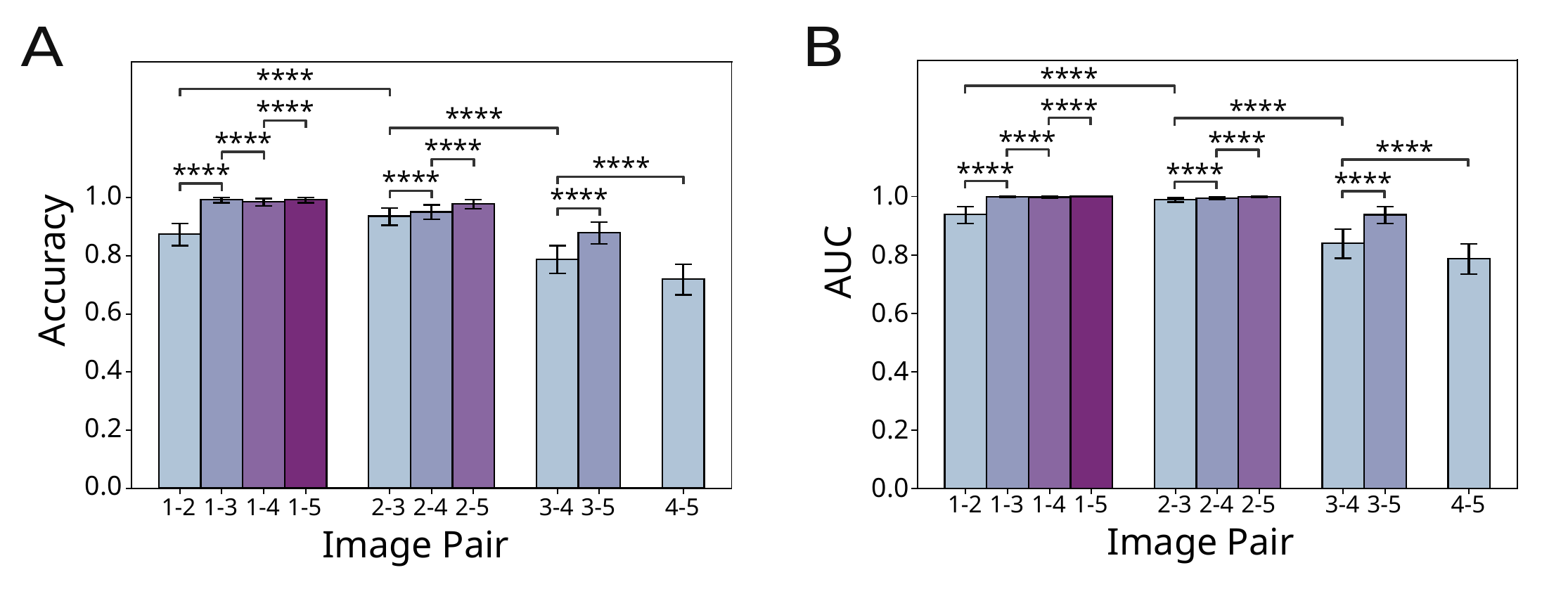}
\end{center} 
\caption{\textbf{Pairwise performance of the \textit{All-pairs} model.}(A) Accuracy and (B) AUC of the \textit{All-pairs} model for different image pairs. Pairs of numbers on the x-axes indicate the fractions at which the paired images were acquired. Cases were grouped by the first fraction and ordered and color-coded by the interval between the paired fractions.
Stars indicate statistical significance between compared cases (independent two-sided t-test; ****: $p \le .0001$).}
\label{fig:figure_score_pair_AB}
\end{figure*}

\clearpage

\begin{figure*}[p]
\begin{center}
\includegraphics[width=0.5\linewidth]{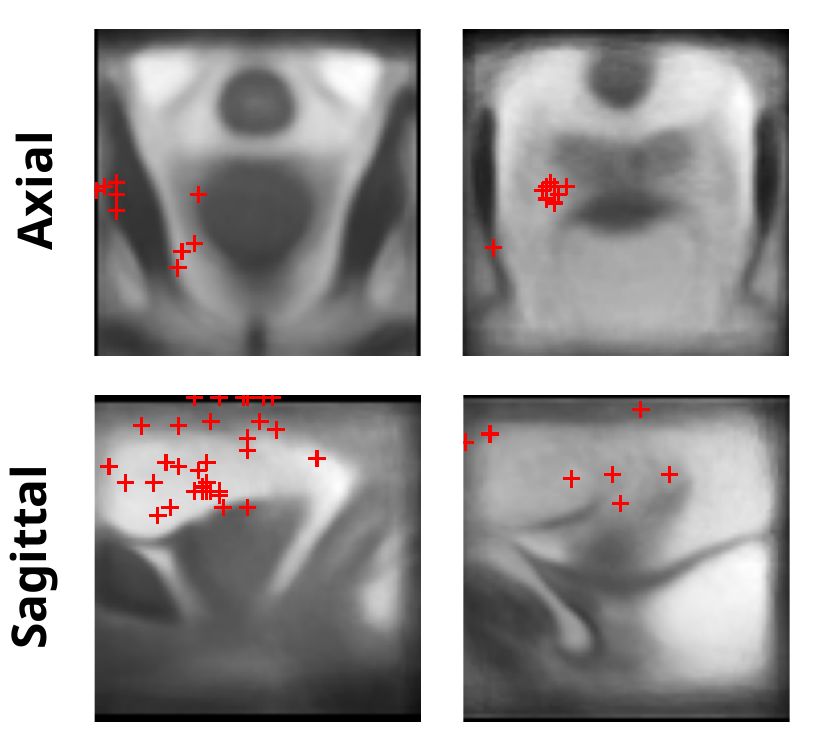}
\end{center} 
\caption{\textbf{Peaks of saliency maps of patients in the test data on MR image atlas.}}
\label{fig:atlas_gradcam_peak}
\end{figure*}

\clearpage
\begin{figure}[p]
    \begin{center}
    \includegraphics[width=0.7\textwidth]{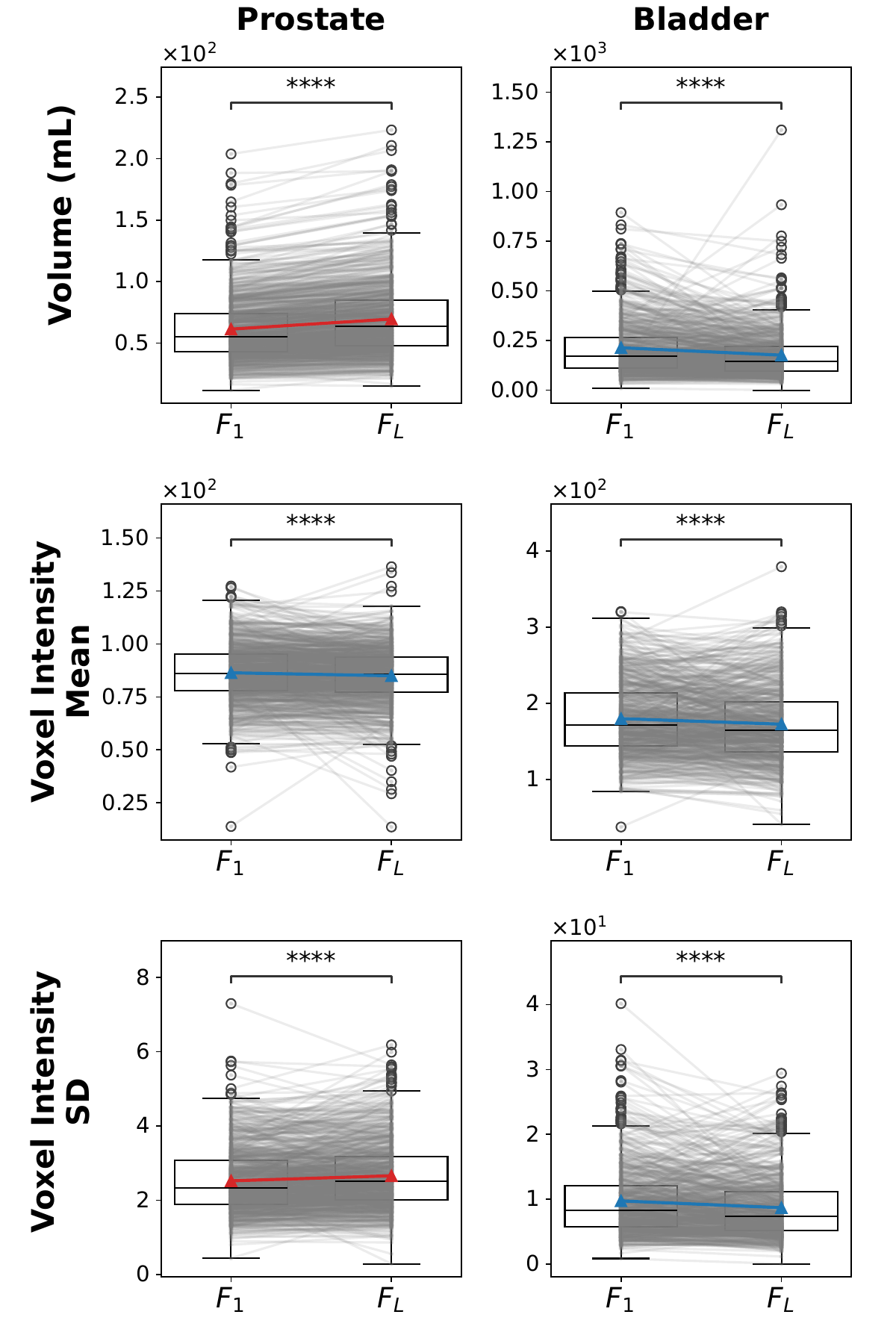}
    \end{center}
    \vspace{-0.5cm}
    \caption{\textbf{Changes in volume and intensity of prostate and bladder in MR images.} An increase in the difference between the two means is indicated in red, whereas a decrease is indicated in blue.
    Stars indicate statistical significance of the difference (Wilcoxon signed-rank test; ****: $p \le .0001$)}
    \vspace{-0.5cm}
    \label{fig:organ}
\end{figure}

\end{document}